\documentclass{elsart}
\usepackage{amssymb}
\usepackage{psfig}

\def\mpunkt{{\hspace{0.5cm} \rm .}}
\def\mkomma{{\hspace{0.5cm} \rm ,} \hspace{0.5cm}}

\begin{document}

\begin{frontmatter}



\title{{\small Chaos, Solitons \& Fractals Vol.5, 1901-1912 (1995)}\\[0.3cm]
Complex Flow of Granular Material in a Rotating Cylinder}


\author{Thorsten P\"oschel$^{*+}$ and Volkhard Buchholtz$^+$}

\address{$\rm ^* $ HLRZ, Forschungszentrum J\"ulich, Postfach 1913,
        D--52425 J\"ulich, Germany  \\ 
        $\rm ^+ $ Humboldt--Universit\"at zu Berlin,
        Institut f\"ur Theoretische Physik,
        Invalidenstra\ss e 110, D--10115 Berlin, Germany}

\begin{abstract}
The flow of granular material in a rotating cylinder was simulated by
molecular dynamics in two dimensions using spherical as well as nonspherical
grains. At very low but constant angular velocity we found that the
flow varies irregularly with time. The particles move stick--slip like i.e.
there are avalanches of different size at the surface of the granular material.
Observing the traces of the particles we found that there are unstable
convection cells. 
Our results agree with recent experiments by Rajchenbach~\cite{rajchenbach} and
Rolf~\cite{vt}.
\end{abstract}

%\begin{keyword}

%\PACS 
%\end{keyword}
\end{frontmatter}

% main text
\section{Introduction}
The flow of granular material in a rotating cylinder is of particular interest
since on the one hand it reveals many interesting to physicists phenomena, 
on the other hand in chemical engineering widely used tumbling ball mills base
on the motion of iron spheres moving in rotating cylinders.
Hence it has been intensively experimentally subjected by both
engineers~\cite{vt} and physicists~\cite{rajchenbach,briscoe}.  
\par
The present paper contains results of two dimensional molecular dynamics of
grains moving in an uniformly rotating cylinder. Recently
Rajchenbach~\cite{rajchenbach} has shown experimentally that the inclination of
sand flowing in a rotating cylinder depends nonlinearly on the angular
velocity of the cylinder. If the angular velocity is smaller than a certain
threshold the sand moves stick--slip like, i.e. there is no continuous flow
but there are avalanches of different size, the angle of inclination
fluctuates. Using high speed recording of polydisperse iron spheres in a
tumbling ball mill Rolf~\cite{vt} observed that the flow forms
convection cells. They found that the size and the number of the cells depend
on the rotation velocity. We will show that our simulations agree
with the experimental results~\cite{rajchenbach,vt,briscoe}. 
\par
Traditional molecular 
dynamics simulations of granular material use polydisperse spherical grains,
examples can be found in~\cite{mdxspherical,cundall,haff} and
many references therein. Applying a recently introduced model for a nonspherical
grain~\cite{poeschelxbuchholtz} which deforms elastically under
pressure and dissipates energy during collisions our results agree better with
the experiment than the simulations using spheres for the case of low angular
velocity. Nonspherical grains in molecular dynamics simulations of
granular material have been introduced already by Gallas and
Soko{\l}owski~\cite{gallasxs} in an investigation of the angle of repose of a 
sand heap. In their model a grain consists of two spheres which are connected
by a stiff bar, their results agree with ours. 

\section{Molecular dynamics of granular media}
In the simulations we assume that the grains are ideal spheres.
To describe the interaction between colliding particles we use a simple Ansatz
given by Cundall and Strack~\cite{cundall} and Haff and Werner \cite{haff}. Two
grains $i$ and $j$ at the positions $\vec{r}_i$ and $\vec{r}_j$ will interact
only if they touch each other, i.e. if
the distance between the center points is smaller than the sum of their radii
$R_i$ and $R_j$
\begin{equation}
\mid \vec{r}_i-\vec{r}_j \mid~<~R_i + R_j \mpunkt
\end{equation}
For this case the force between particles $i$ and $j$ moving with the velocities
$\dot{\vec{r}}_i$ and $\dot{\vec{r}}_j$ and rotating with angular velocities
$\dot{\Omega}_i$ and $\dot{\Omega}_j$ is given by 
\begin{equation}
\vec{F}_{ij} =  F_{ij}^N \cdot \frac{\vec{r}_i
-\vec{r}_j}{|\vec{r}_i-\vec{r}_j|} + F_{ij}^S \cdot  \left({0 \atop 1} ~{-1
\atop 0} \right) \cdot \frac{\vec{r}_i - \vec{r}_j}{|\vec{r_i}-\vec{r}_j|}
\mpunkt 
\end{equation}
with the normal force
\begin{equation}
        F_{ij}^N  = Y \cdot (R_i + R_j - |\vec{r}_i - \vec{r}_j|)^{\frac{3}{2}}+
        ~\gamma_N \cdot m_{ij}^{eff}\cdot |\dot{\vec{r}}_i - \dot{\vec{r}}_j|
\end{equation}
and the shear force
\begin{equation}
        F_{ij}^S = \min \{- \gamma_S \cdot m_{ij}^{eff} \cdot |\vec{v}_{ij}^{rel}|~,~ \mu \cdot 
        |F_{ij}^N| \} \mpunkt
\label{eq_coulomb}      
\end{equation}
For the relative velocity $\vec{v}_{ij}^{rel}$ of the particle surfaces one can write
\begin{equation}
        \vec{v}_{ij}^{rel} = (\dot{\vec{r}}_i - \dot{\vec{r}}_j) + R_i \cdot
\dot{\Omega}_i + R_j \cdot \dot{\Omega}_j \mpunkt
\end{equation} 
The effective mass of the particles $i$ and $j$ is given by
\begin{equation}
m_{ij}^{eff} = \frac{m_i \cdot m_j}{m_i + m_j} \mkomma
\label{eq_eff_mass}
\end{equation}
and the resulting momenta $M_i$ and $M_i$ for the particles $i$ and $j$ are
\begin{equation}
M_i = F_{ij}^S \cdot R_i
\end{equation}
and
\begin{equation}
M_j = - F_{ij}^S \cdot R_j \mpunkt
\end{equation}
$Y$ is the Young modulus, $\gamma_N$ and $\gamma_S$ are the phenomenological
normal and shear friction coefficients, $\mu$ stands for the Coulomb
friction parameter.
\par
The underlying assumption is that the particles deform each other slightly
during collisions. To simulate three dimensional behavior we use the Hertzian
contact force \cite{landau} which rises with 
the power 1.5 of the penetration depth $R_i + R_j - |\vec{r}_i - \vec{r}_j|$.
Eq.~(\ref{eq_coulomb}) takes into account that the particles slide on each
other for the case that the inequality
\begin{equation}
\mu \cdot \mid F_{ij}^N \mid~<~\mid F_{ij}^S \mid
\end{equation}
holds, otherwise they roll. This behavior is due to the Coulomb law.
\par
For the integration of Newtons equations of motion due to the forces
$\vec{F}_{ij}$ given above we used a sixth order Gear predictor--corrector
algorithm~\cite{predictor}. This high accuracy is necessary since there is a
short range interaction between the particles, a small inaccuracy in the
calculation of the particle positions may cause large faults in the resulting
forces. Using algorithms of lower accuracy the
system would become numerical unstable. The calculation of the rotation of the
particles is not such sensitive since the angular velocity is very small in
general, hence we applied a forth order algorithm only. 
\par
In the simulations we use for the parameters $Y=10^4 kg/s^2$,
$\gamma_N=1.5\cdot 10^4~s^{-1}$, $\gamma_S=3\cdot 10^{4}~s^{-1}$, $\mu
=0.5$.  The radii of the particles are Gauss distributed with the mean
value $\overline{R_i}=1~mm$. The chosen parameters are typical for a
soft granular material. 
%\section{Results}
\par
\begin{figure}[htbp]
\centerline{\psfig{figure=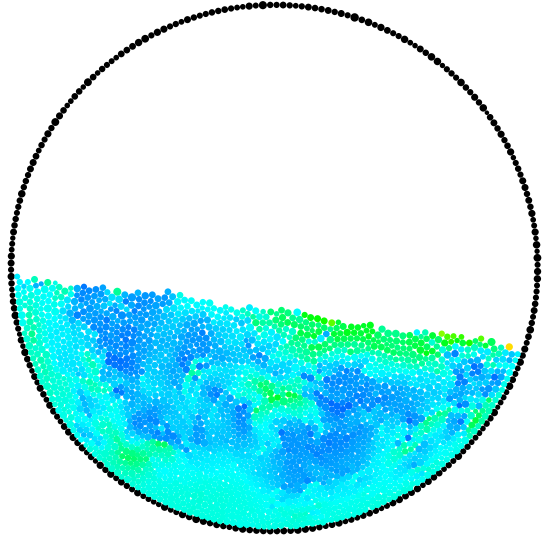,width=7cm}}
\caption{Snapshot of the simulation of 2400 spherical grains moving in a
cylinder which rotates with constant angular velocity $\Omega=0.002~sec^{-1}$.
High velocity is coded by red color, particles moving with low velocity are
drawn blue.}
\label{snap_sph0.002}
\end{figure}

The model described above was applied to a simulation of granular
material consisting of 2400 single grains moving in a clockwise
rotating cylinder of diameter $160\cdot \overline{R_i}$, the cylinder
rotates with angular velocity $\Omega=0.002~sec^{-1}$.
Fig.~\ref{snap_sph0.002} shows a snapshot of the simulation. 

\begin{figure}[htbp]
\centerline{\psfig{figure=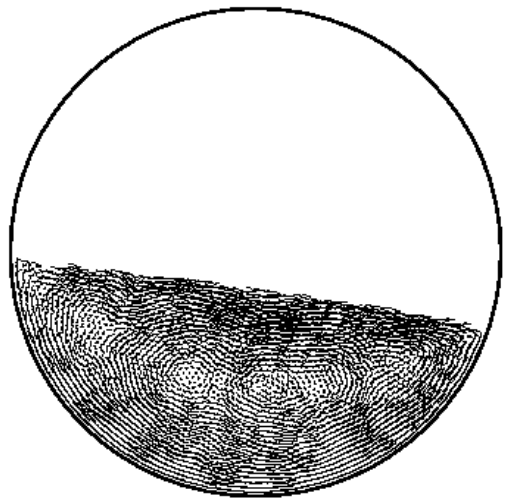,width=7cm}}
\caption{Traces of the motion of granular material at an angular velocity
$\Omega=0.002~sec^{-1}$. One observes several unstable convection cells.}
\label{strom_0.002}
\end{figure}

The color
codes for the velocity of the particles, slow particles are drawn blue,
fast are drawn red. As visible the velocity of the particles near the
surface of the flow is nonuniform.  The surface of the granular
material is close to an even plane which is inclined by an angle
$\Theta \approx 8^o$. Although the dynamics of the flow seems to be
very regular the traces of the grains during the simulation are due to
convection cells~(fig.\ref{strom_0.002}). Since there are more than
one cell they cannot remain stable over a long time because all of
them have the same sense of rotation and therefore they disturb each
other. This effect was observed experimentally too in the
investigation of tumbling ball
mills~\cite{vt}. 

\begin{figure}
\centerline{\psfig{figure=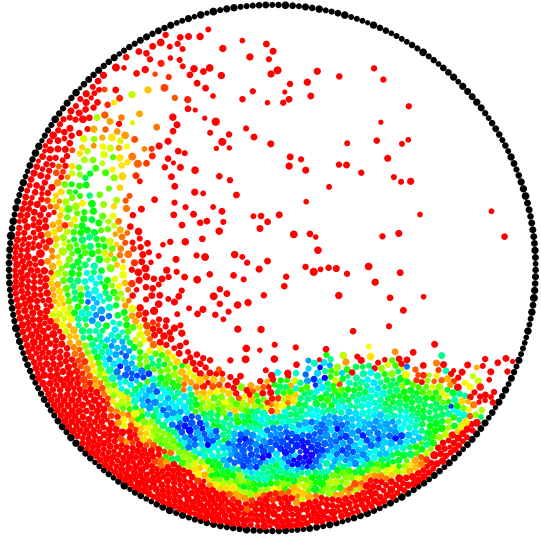,width=7cm}}
\caption{Snapshot of the simulation of 2400 spherical grains moving in a
cylinder which rotates with high angular velocity $\Omega=0.1~sec^{-1}$. The
bulk of the granular material is S--shaped.}
\label{snap_sph0.1}
\end{figure}
\begin{figure}
\centerline{\psfig{figure=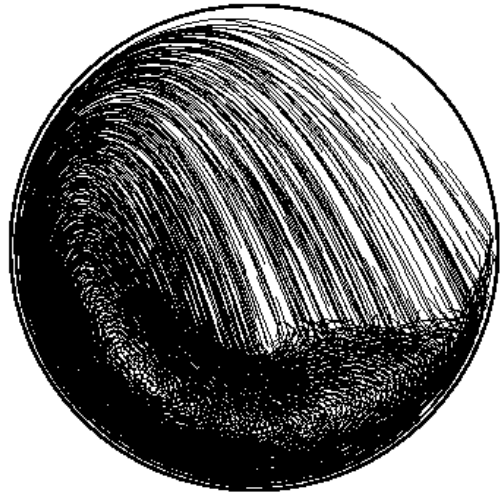,width=7cm}}
\caption{Traces of the particles for a high angular velocity
($\Omega=0.1~sec^{-1}$). There is only one large convection cell.}
\label{strom_0.1}
\end{figure}

%snap und strom schnell *****************************************************
Figs.~\ref{snap_sph0.1} and \ref{strom_0.1} show a snapshot and the plot
of the traces of a simulation with angular velocity $\Omega=0.1~sec^{-1}$. The
surface of the granular material is not an even plane anymore, but it
becomes S--shaped. The same shape characteristic was observed in
experiments by Rajchenbach~\cite{rajchenbach}.
In the simulations for high angular velocity one finds only one large
convection cell (fig.~\ref{strom_0.1}) instead of several for lower angular
velocity, this behavior is due to the experiment~\cite{vt}.

After submitting this
article we received a preprint by Baumann, Jobs and Wolf~\cite{wolf} who applied
the interesting simulation method by Visscher and
Bolsterli~\cite{visscher} (improved version~\cite{jullien}) to the
problem of the rotating cylinder. They found convection cells as well.
\par
\begin{figure}
\centerline{\psfig{figure=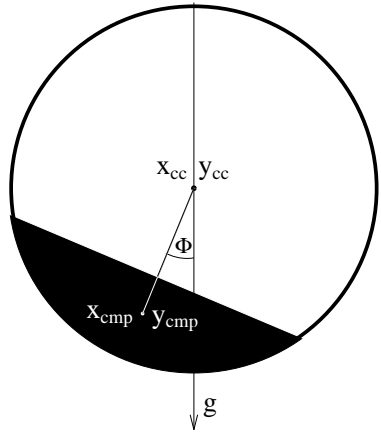,width=5cm,angle=270}}
\caption{Schematic description of the procedure we used to determine the
inclination $\Theta$.}
\label{w_m}
\end{figure}

%winkelmessung **************************************************************
Because the surface of the simulated granular flow is not a perfect plane even
for very slow rotation we
have to determine the inclination indirectly. It is given by
\begin{equation}
\Theta=\arctan \left( \frac{x_{cc}-x_{cmp}}{y_{cc}-y_{cmp}}\right),
\end{equation}
where ($x_{cmp}$~,~$y_{cmp}$) are the coordinates of the center of mass point
and ($x_{cc}$~,~$y_{cc}$) is the rotation center of the cylinder
(fig.~\ref{w_m}).

\begin{figure}
\centerline{\psfig{figure=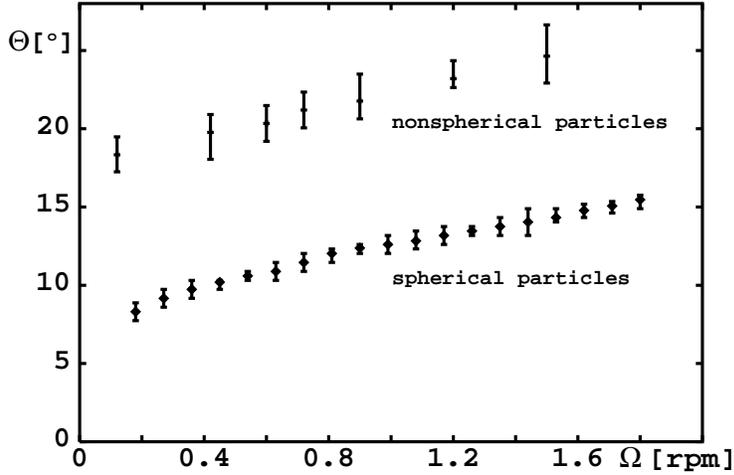,width=10cm,angle=270}}
\caption{Inclination $\Theta$ as a function of the angular velocity
$\Omega$ of the cylinder for spherical (a) and nonspherical (b) grains. In both
cases the inclination rises as $\Theta-\Theta_c \sim \Omega^{0.8}$, where
$\Theta_c \approx 8^o$ for spheres and $\Theta_c \approx 18^o$ for nonspherical particles.}
\label{winkel_omega}
\end{figure}

%winkel omega ***************************************************************
As shown in figs.~\ref{snap_sph0.002} and \ref{snap_sph0.1} the profile of the
granular material depends on the angular velocity. 
Curve (a) in fig.~\ref{winkel_omega} shows the relation between the angular
velocity 
$\Omega$ and the inclination $\Theta$. It rises
almost linear with the angular velocity. Using the Ansatz 
\begin{equation}
\Theta-\Theta_c \sim \Omega^m
\end{equation}
we found for the exponent $m \approx 0.8$ for the interval $0<\Omega<2~rpm$.
For larger angular velocities the profile is not close to an even plane but
S--shaped. 

\begin{figure}[htbp]
%\centerline{\psfig{figure=fluss.sphxxx.ps,height=22cm}~~~\psfig{figure=fluss.geoxxx.ps,height=22cm}}
\centerline{\psfig{figure=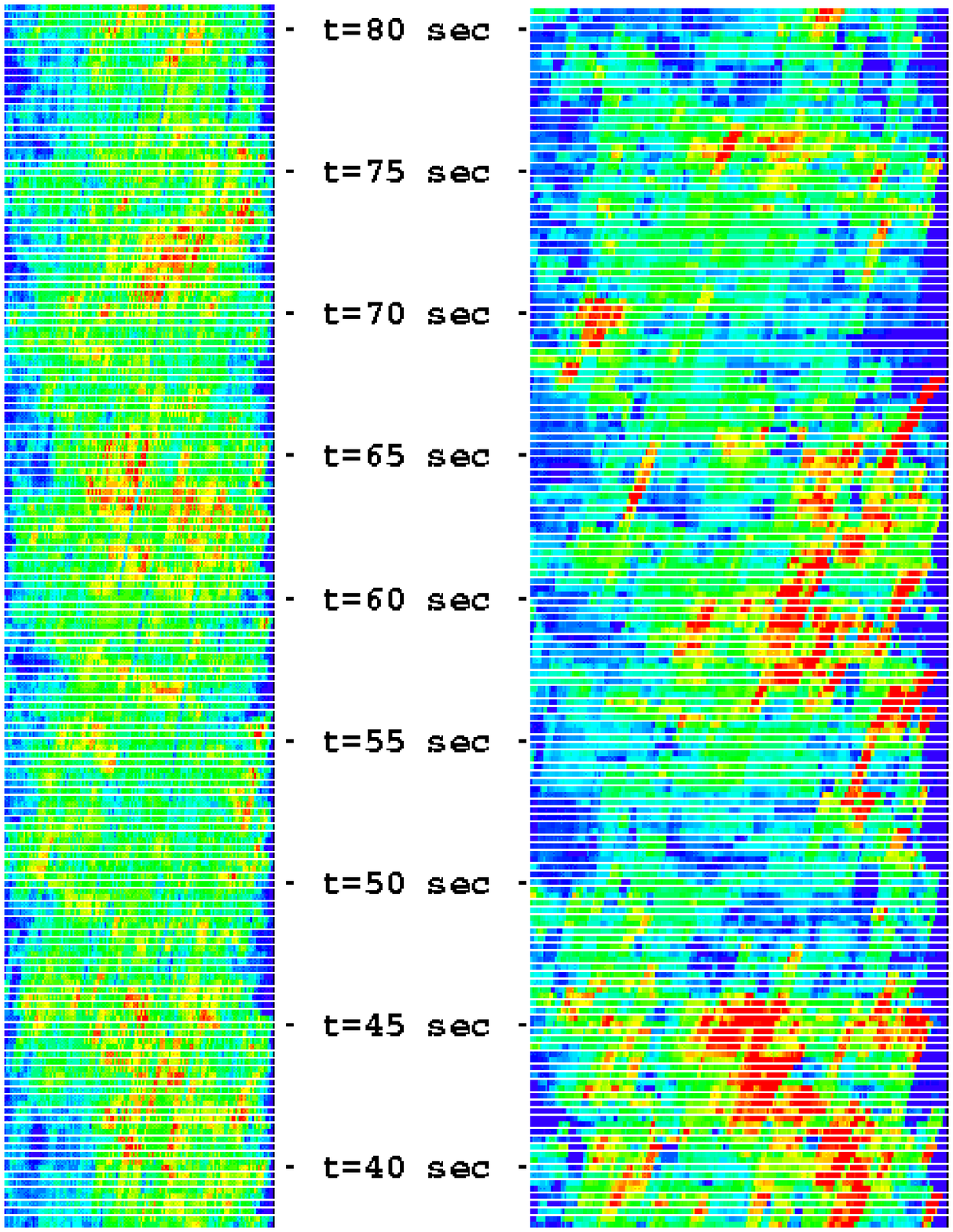,height=18cm}}
\caption{(left) Evolution of the velocity of the particles moving at the surface
of the flow. Each raw is a snapshot. The flow varies irregularly in time and
space.}
\label{fluss_sph}
\end{figure}

\begin{figure}
\centerline{\psfig{figure=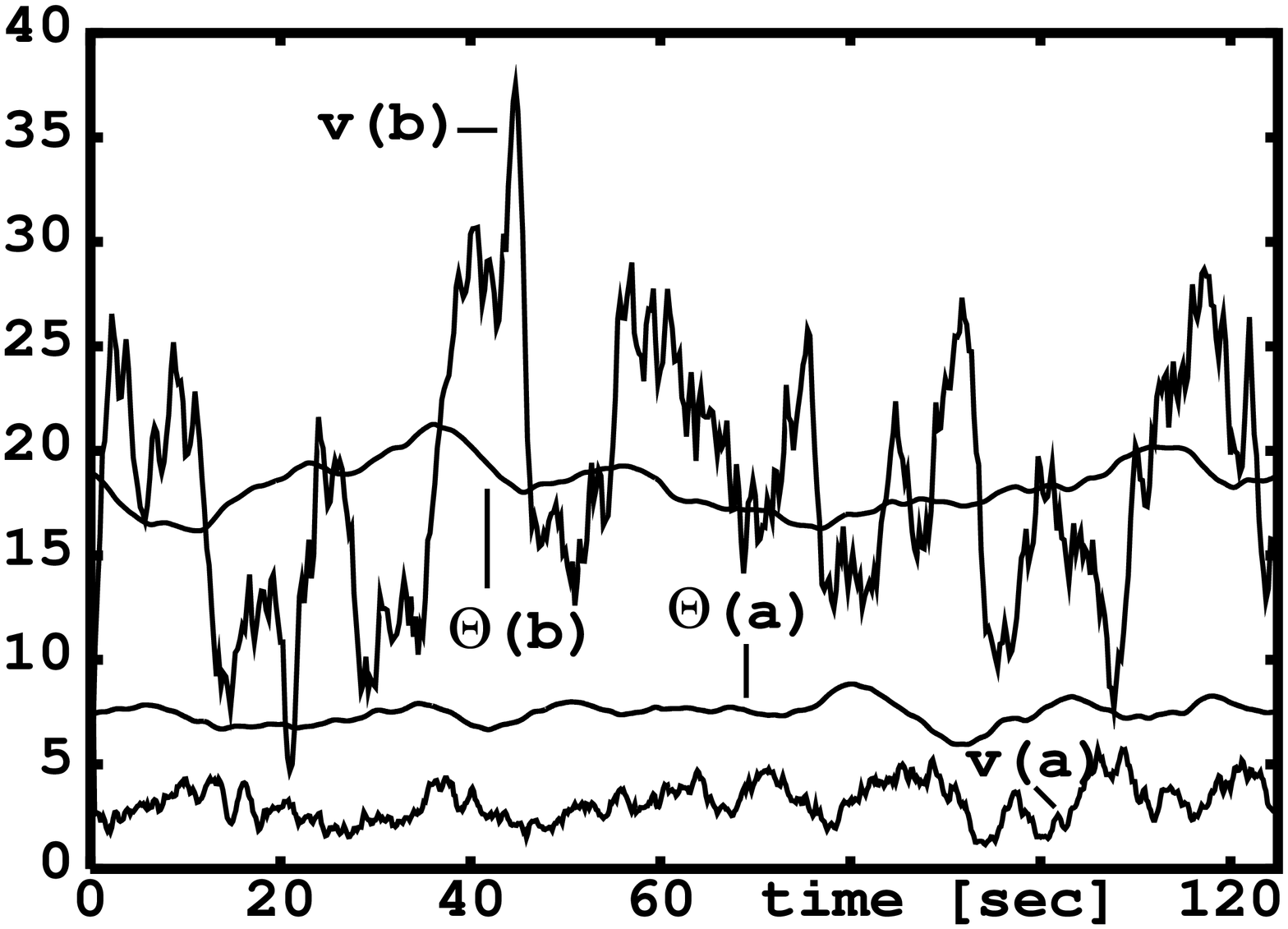,width=10cm,angle=270}}
\caption{The inclination $\Theta [^O]$ and the average velocity
$\overline{v} [50~sec^{-1}]$
of the surface particles for spherical (a) and nonspherical grains (b). While
$\overline{v(b)}$ fluctuates drastically due to avalanches $\overline{v(a)}$ is
much smoother.}
\label{winkel_v_t}
\end{figure}

%fluss***********************************************************************
Although the cylinder rotates with a constant angular velocity the flow of the
surface particles varies irregularly with time and space, this behavior is
shown in fig.~\ref{fluss_sph}. Each horizontal strip of this figure shows the
velocity distribution along the surface of the flow. As before red color means
high velocity, blue means low velocity. The irregular flow was experimentally
observed by Rajchenbach~\cite{rajchenbach} and Briscoe~et.~al.~\cite{briscoe}
before. Curves (a) in fig.~\ref{winkel_v_t} show the time evolution of the
inclination and of the particle velocity at the surface, both values fluctuate
irregularly. 
\begin{figure}
%\centerline{\psfig{figure=fluss.geoxxx.ps,height=22cm}}
\caption{(right) Flow at the surface of the granular material in a simulation using
nonspherical grains. There are successive regions of higher and lower energy
due to avalanches. The color contrasts are sharper than in fig. 7.}
\label{fluss_geo}
\end{figure}

\section{A more complex model}
In the previous section we have shown that the granular flow is irregular, however,
fig.~\ref{winkel_v_t} proves that there are no sharp bounded avalanches.
Experimental observations, however, show that the flow at the surface of granular
material in a rotating cylinder is stick--slip like, i.e. there are avalanches
of different size going down the surface. Our simulations with spherical grains
cannot reproduce this behavior. Hence we apply another model where the grains
are not simple spheres but of more complex shape.
\par
\begin{figure}
\centerline{\psfig{figure=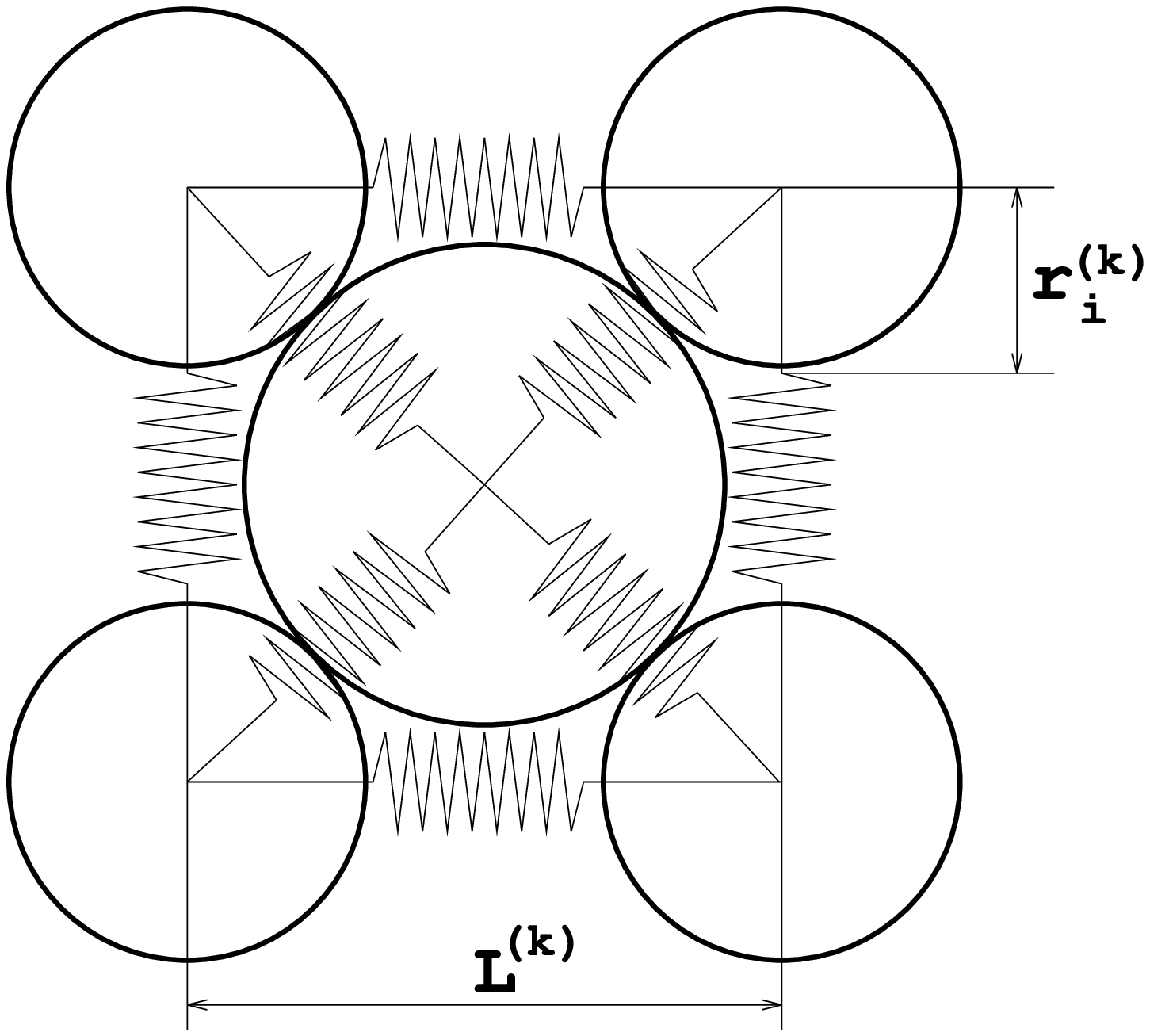,width=5cm}}
\caption{The nonspherical particles consist of four spheres at the corners
of a square and one in the middle of the square which radius is chosen to touch
the others.}
\label{geopart}
\end{figure}

%erklaerung schema ********************************************************
Each nonspherical particle $k$ consists of four spheres of radius $r_i^{(k)}$ located
at the corners of  a square of size $L^{(k)}$ and one additional sphere in the
center of the square (fig.~\ref{geopart}). The radius $r_m^{(k)}$ of the middle
sphere is chosen to touch the others 
\begin{equation}
r_{m}^{(k)} = L^{(k)} / \sqrt{2}- r_{i}^{(k)} \mpunkt
\end{equation}
During each collision between to spheres $i$ and $j$ with radii $R_i$ and $R_j$
which might belong to the same grain or to different grains acts a force in normal
direction 
\begin{equation}
\vec{F}_{ij}^{C} = \Big[ Y (R_i+R_j-|\vec{r}_i-\vec{r}_j|)+\gamma_N
         m_{ij}^{eff} |\dot{\vec{r}}_i - \dot{\vec{r}}_j| \Big]
        \frac{\vec{r}_i - \vec{r}_j}{|\vec{r_i}-\vec{r}_j|}~,
\end{equation}
where the effective mass is given by eq.~(\ref{eq_eff_mass}). The notation is
the same as above.
In addition each pair of neighboring spheres $i$ and $j$ which both belong to the same
grain feel a force $\vec{F}_{ij}^{Sp}$ due to a damped spring
\begin{equation}
\vec{F}_{ij}^{Sp} = \Big[ \alpha (C^{(k)} - |\vec{r}_i-\vec{r}_j|) + 
        \gamma_{Sp}\frac{m_i}{2} |\dot{\vec{r}}_i -
\dot{\vec{r}}_j | 
        \Big]
        \frac{\vec{r}_i - \vec{r}_j}{|\vec{r}_i-\vec{r}_j|}~,
\end{equation}
$\alpha$ stands for the spring constant and $\gamma$ is the damping coefficient.
The term $C^{(k)}$ is the length of the spring in equilibrium
\begin{equation}
C^{(k)}_{ij} = \left\{ \begin{array}{ll}
                L^{(k)} & \mbox{if $i$, $j$ lie at the same edge}\\
                L^{(k)}/\sqrt{2} & \mbox{else} \mpunkt
                \end{array} \right.
\end{equation}
We want to point out here that eq.~(\ref{eq_coulomb}) which introduces static
friction into the interaction forces is not applied in the simulations using
nonspherical grains.
\par
We investigated the flow of 1000 nonspherical grains, where the parameters
are the same as in the previous simulation and 
$\alpha=10^4~kg/s^2$, $\gamma_{Sp}=3\cdot 10^4~s^{-1}$, $L^{(k)}/r_i^{(k)}=4$
and $\overline{L^{(k)}}=3~mm$. The cylinder of diameter $D\approx 86\cdot
\overline{L^{(k)}}$ 
rotates with angular velocity $\Omega=0.002~sec^{-1}$.
\par
\begin{figure}
\centerline{\psfig{figure=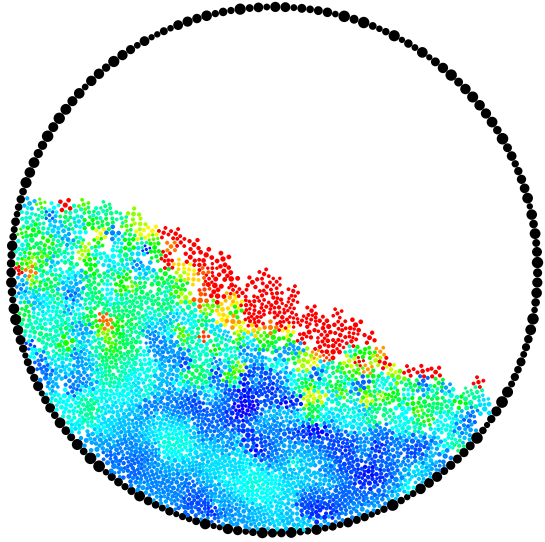,width=7cm}}
\caption{Snapshot of the simulation of 1000 nonspherical grains each
consisting of 5 spheres moving in a
cylinder which rotates with constant angular velocity $\Omega=0.002~sec^{-1}$.
The red drawn grains at the surface show an avalanche.}
\label{snap_geo0.002}
\end{figure}

%snap ********************************************************************
Fig.~\ref{snap_geo0.002} shows a snapshot of the simulation. Obviously the
inclination is approximately twice the inclination observed for
spherical grains. The particles at the surface move with much higher velocities
than the others. This is due to an avalanche going down the surface in the
moment when the snapshot was done. 

%streifen ***************************************************************
The equivalent picture to fig.~\ref{fluss_sph} is shown in
fig.~\ref{fluss_geo}. The flow is irregular in time and space too but one
finds time intervals where the surface particles move with high velocities and
other intervals where the particle velocities, i.e the flow at the 
surface, are almost zero. This succession indexes stick--slip motion. The time
evolution of the average velocity $\overline{v}$ of the surface particles and
of the inclination $\Theta$ of the surface for nonspherical (curves(b)) as well
as for spherical grains (curves(a)) is drawn in fig.~\ref{winkel_v_t}. While in
the case of spherical grains $\overline{v}$ varies only slightly for the case
of nonspherical particles there are drastic fluctuations of the average
velocity due to avalanches of different size.

%winkel ******************************************************************
For the model of nonspherical particles we investigated the dependence of the
angle $\Theta$ of the surface on the angular velocity $\Omega$ too
(fig.~\ref{winkel_omega}). The measured inclinations are approximately
twice the inclinations observed in simulations with spheres. These values agree
better with experimental investigations. For the exponent $m$ in
$\Theta-\Theta_c \sim \Omega^m$ we found $m\approx 0.8$.

\section{Conclusion}
The present paper reports results of two dimensional molecular dynamics
simulations of granular material moving in a cylinder which rotates with 
constant angular velocity. Using spherical particles and the Ansatz of Cundall and
Strack \cite{cundall} for the simulation of static friction we found that the
flow fluctuates 
irregularly with time. The angle $\Theta$ of the material as well as the
velocity of the grains at the 
surface fluctuate significantly. For the dependence of the inclination $\Theta$
of the granular material on the angular velocity $\Omega$ of the cylinder
we found $\Theta-\Theta_c \sim \Omega^{0.8}$. The flow of the particles forms
unstable convection cells. For low rotation velocity we observed several cells,
for high 
rotation velocity we observed only one cell. In the latter case the bulk of the
granular flow reveals a typical S--shape.
\par
When using nonspherical grains, where each grain consists of four spheres
located at the corners of a square and a fifth sphere at the center, and where
neighboring spheres of the grain are connected by a damped spring, we found
that the flow moves stick--slip like. That means there are avalanches of
different size at the surface of the flow. The inclination of the granular
material $\Theta$ is more than two times larger than in the simulations using
spheres which agrees better with the experiment. 
\par
In experiments was observed that for low angular velocity $\Omega$ the
particles move stick--slip like due to avalanches going down the inclined
surface, for higher velocity one found continuous flow and the characteristic
S--shape~\cite{rajchenbach}.  The inclination depends nonlinearly on the
angular velocity. In experiments with tumbling ball mills unstable
convection cells have been observed~\cite{vt}.
\par
Our simulations reproduced the experimentally measured effects using simple
assumptions about the behavior of a grain during the collisions. The dynamic
effects as irregular flow at low rotation velocity, the finite inclination and
the existence of convection cells could be reproduced using simple spherical
grains. To simulate the avalanches at the surface of the flow, which is an
effect of the static friction between the grains, we applied a more complex
model for grains which includes nonsphericity. \vspace{1cm}\\

\section*{Acknowledgment}
The authors thank H.~J.~Herrmann for helpful discussions.

\end{document}